\documentclass[11pt]{article}
\usepackage[preprint]{acl}
\usepackage{multirow}

\usepackage{times}
\usepackage{latexsym}

\usepackage[utf8]{inputenc} 
\usepackage[T1]{fontenc}    
\usepackage{hyperref}       
\usepackage{url}            
\usepackage{booktabs}       
\usepackage{amsfonts}       
\usepackage{nicefrac}       
\usepackage{microtype}      
\usepackage{xcolor}         
\usepackage{wrapfig}
\usepackage{graphicx}
\usepackage{amsmath} 
\usepackage{comment}       
\usepackage{subcaption}
\usepackage{tcolorbox}
\tcbuselibrary{skins}
\usepackage{microtype}
\usepackage{inconsolata}

\definecolor{gtbg}{HTML}{F8FAFC}
\definecolor{addc}{HTML}{1B5E20}
\definecolor{delc}{HTML}{B71C1C}
\definecolor{easyc}{HTML}{2E7D32}
\definecolor{medc}{HTML}{EF6C00}
\definecolor{hardc}{HTML}{C62828}

\title{\texttt{TeleSWEBench}:  A Commit-Driven Benchmark for Evaluating LLM-Powered Software Engineering in Telecommunications}

\author{%
  Pranshav Gajjar\\
  NCSU, USA\\
  \And
  Ali Mamaghani\\
  UC San Diego, USA\\
  \And
  Dinesh Bharadia\\
  UC San Diego, USA\\
  \And
  Vijay K Shah\\
  NCSU, USA\\
}

\begin{document}

\maketitle

\begin{abstract}
With the telecommunications field embracing \textit{zero touch} management alongside novel O-RAN and AI-RAN frameworks, contemporary telecom networks now function as immensely intricate and heavily softwareized codebases. While automated software engineering (ASE) tools and Software Engineering (SWE) \textit{Agents} hold the potential to alleviate the critical code generation bottleneck in this domain, their ability to navigate and modify specialized, mathematically rigorous wireless stacks like \texttt{srsRAN 5G} remains unverified. General-purpose coding benchmarks fail to capture the stateful logic and strict requirements of telecommunications, leaving a critical evaluation gap. In this paper, we introduce \textbf{TeleSWEBench}, the first commit-driven benchmark specifically designed to measure an agent's performance in the telecom domain. We mine real developer commits from the \texttt{srsRAN 5G} repository and distill them into structured test cases across three difficulty tiers (Easy, Medium, and Difficult). Our benchmark consists of 
734 questions that are accompanied by executable unit tests. To avoid the rigidity of test cases, we further propose a hierarchical LLM as a Judge framework called \textit{TeleJudge} that scores agent outputs at the file level and aggregates verdicts holistically. This follows an evaluation based on context and semantic similarity in parallel to a standard unit test-based evaluation. Using this benchmark, we evaluate AIDER, OpenHands, and the ClaudeCode frameworks, powered by state-of-the-art reasoning LLMs, including Qwen3, GPT OSS, Gemma 4, Kimi, and Qwencoder 2.5. Our two-stage evaluation reveals that models suffer from a lack of both localization accuracy and functional correctness, with the strongest ASE tools achieving up to 25\% of shippable changes.
\end{abstract}

\section{Introduction}

The telecommunications industry is undergoing a fundamental paradigm shift toward fully autonomous, \textit{zero touch} network provisioning and management. Driven by the advent of Open RAN (O-RAN) \cite{tripathi2025fundamentals} and the emerging Artificial Intelligence-Radio Access Network (AI RAN) \cite{FENG20265} architectures, traditional proprietary telecom hardware is being rapidly supplanted by highly softwarized, cloud native network functions \cite{tripathi2025fundamentals}. Consequently, modern wireless networks, from the physical layer up to the core, are essentially massive, highly complex codebases deployed on commercial off-the-shelf hardware \cite{polese2023understanding}. As the demand for dynamic resource allocation, continuous protocol upgrades, and rapid feature deployment accelerates, the primary bottleneck in realizing these next-generation networks has shifted from hardware limitations to automated software engineering and code generation.

Concurrently, the rapid advancement of Large Language Models (LLMs) has catalyzed a revolution in automated software engineering (ASE). Autonomous \textit{SWE} agents can now navigate codebases, generate multi-file patches, and resolve standard GitHub issues with increasing autonomy \cite{hou2024large, zhang2026survey}. These tools promise to alleviate the critical software bottleneck in telecom by automating the maintenance and expansion of network stacks. However, the efficacy of these general-purpose agents remains largely unverified when applied to the mathematically rigorous, highly specialized, and hardware adjacent domain of wireless communications. Wireless software stacks, such as the widely adopted open source \texttt{srsRAN 5G} \cite{Paisana2026srsran, gajjar2025oransight}, impose unique and unforgiving constraints: strict asynchronous timing requirements, intricate state machine management, layered protocol hierarchies, and rigid adherence to complex 3GPP standardization documents \cite{baron2018unpacking, ganiyu2025ai5gtest}.

Despite the proliferation of these automated coding tools, the research community lacks a systematic methodology to evaluate their true utility in this domain. Existing software engineering benchmarks, such as SWE Bench \cite{jimenez2023swe}, MBPP \cite{austin2021program}, and HumanEval \cite{chen2021evaluating}, evaluate models primarily on generic web development tasks, standard algorithms, or popular Python libraries. They fundamentally fail to capture the domain-specific nuances and cascading cross-file dependencies inherent to a 5G network stack. Existing instruction tuning evaluations, when applied to telecom, often produce shallow or factually incorrect questions. 

As a result, we are left with a critical evaluation gap:  \textit{While we have access to powerful automated software coding tools, there is currently no quantitative way to know which models actually work for softwarized network generation, nor how they fail when confronted with complex telecom logic}. To bridge this gap, we introduce \textbf{TeleSWEBench}\footnote{The benchmarking suite, along with the associated code, is available at \url{https://github.com/prnshv/TeleSWEBench}.}, the first commit-driven benchmark specifically designed to evaluate LLM-powered ASE frameworks and SWE Agents in the telecommunications domain. Rather than relying on synthetic or instruction-tuned queries, TeleSWEBench is constructed by mining real-world developer commits from the \texttt{srsRAN 5G}\footnote{The leading open source O-RAN 5G solution from SRS with an AGPL-3.0 license.} repository \cite{Paisana2026srsran}, distilling authentic bug fixes, feature additions, and protocol updates into structured test cases across three difficulty tiers. We establish a rigorous, two-stage evaluation pipeline that decouples the capability of an agent to navigate a complex repository from its ability to write factually correct domain logic. By structuring the benchmark this way, we can measure scope localization entirely independently of functional correctness. Furthermore, we propose a novel hierarchical LLM as a Judge framework called \texttt{TeleJudge}, which is rigorously validated against repository native executable unit tests to accurately score multiple file patches. Through extensive evaluation of state-of-the-art models within the AIDER \cite{paul2026Aider} and OpenHands \cite{wang2025openhands} frameworks, we expose a significant capability gap in modern LLMs regarding domain-specific code generation. Ultimately, this benchmark provides a critical foundation for the future development of telecom native SWE agents.

\section{Related Work}

\begin{figure}
  \includegraphics[width=0.43\textwidth]{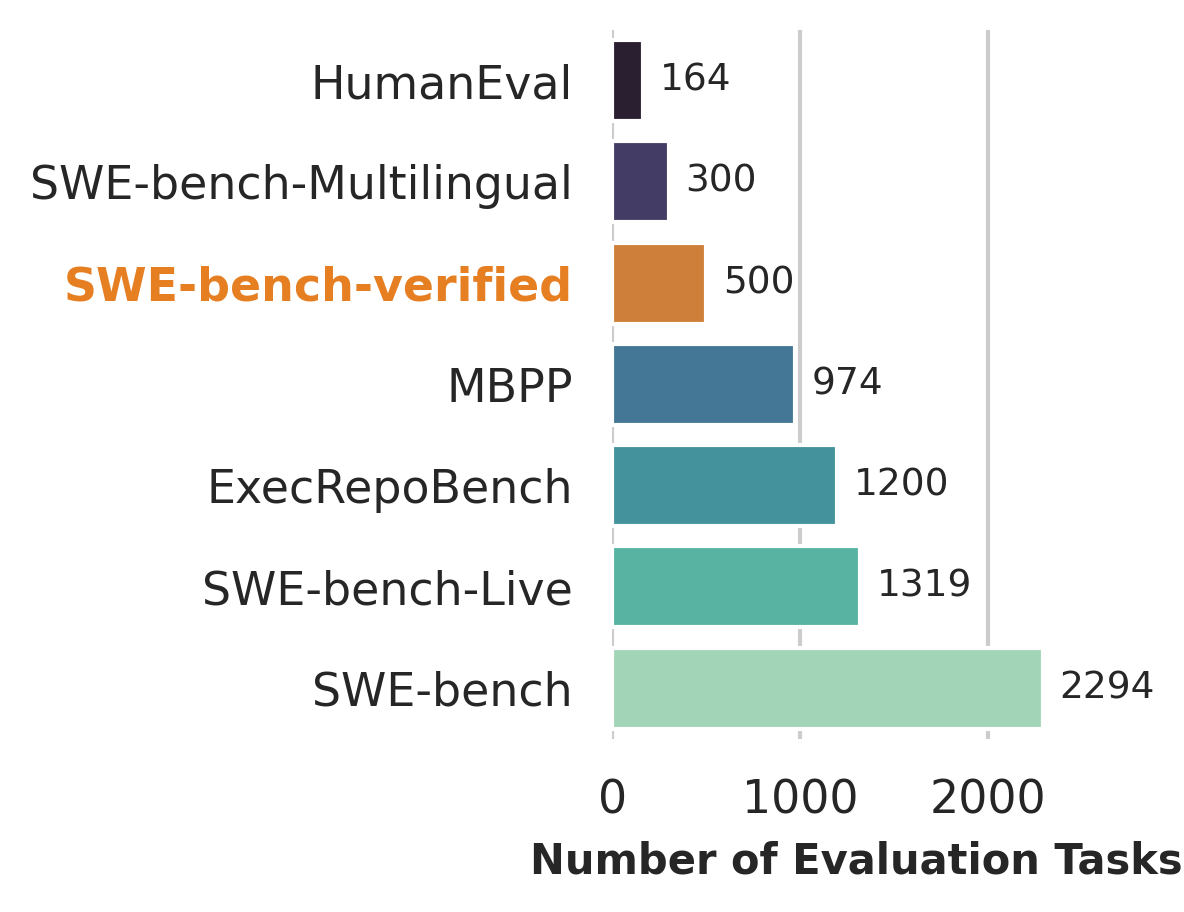}
  \caption{Comparison of the total number of evaluation tasks across prominent Software Engineering (SWE) benchmarks.}
  \label{fig:swe_benchmarks}
\end{figure}

Recent years have seen a proliferation of benchmarks designed to evaluate the coding capabilities of Large Language Models (LLMs). Early datasets, such as HumanEval \cite{chen2021evaluating}, focused on isolated, single-function Python generation tasks, evaluating models on \textit{164} hand-crafted programming problems. As model capabilities advanced, the focus shifted from standalone function synthesis to repository-scale software engineering. SWE-bench \cite{jimenez2023swe} introduced a paradigm shift by requiring models to resolve real-world GitHub issues by navigating multi-file codebases and generating functional patches, offering \textit{2,294} evaluation instances. Swe-Bench has been further extended to other benchmarking suites like \cite{yang2024swe, zhang2025swe}. Similarly, benchmarks such as RepoBench \cite{liu2023repobench} target repository-level code auto-completion and context retrieval, with execution-based variants such as ExecRepoBench \cite{yang2024execrepobench} that evaluate \textit{1,200} repository-level unit tests. The current gold standard for evaluation is SWE-Bench-Verified \cite{chowdhury2024swebenchverified}, which comprises \textit{500} samples verified by experts, and all prominent benchmarks, along with their scale, are shown in Figure \ref{fig:swe_benchmarks}.

While these SWE benchmarks provide excellent testbeds for general-purpose programming and web development logic, they predominantly rely on Python or Java environments. They fundamentally lack the hardware-adjacent, stateful, and mathematically complex C++ paradigms required in telecommunications. Another important aspect is evaluating the multi-file patches generated by autonomous agents. While execution-based evaluation, like compiling and running unit tests, remains the gold standard \cite{yang2024swe, chowdhury2024swebenchverified}, it becomes inherently difficult to evaluate domains or code repositories where \textbf{no} public unit tests are available. Consequently, the use of strong LLMs as judges \cite{gu2024survey,li2024llms,li2025generation} has become a standard practice for evaluating generative LLM outputs for multiple domains, including software development \cite{he2026llm}. The core intuition here is that finding problems in anything is a \textit{significantly} easier task than creating something new, hence using a model that is only prompted to find issues with targeted instructions is capable of reaching an Expert level performance in evaluation. 
Despite rapid advancements in general software engineering, the application and evaluation of LLMs within the telecommunications domain have historically been restricted to Natural Language Processing (NLP) and static information retrieval. Consequently, existing telecom evaluation frameworks are overwhelmingly confined to Multiple-Choice Question Answering (MCQA) formats. Benchmarks such as ORANBench \cite{gajjar2025oran} and TeleQnA \cite{maatouk2025teleqna}, alongside holistic evaluation suites introduced by the GSMA \cite{gsma_release_2026}, are strictly designed to test a model's ability to parse complex 3GPP standardization documents, summarize 5G architectures, or answer theoretical protocol queries. When it comes to actual codebase evaluation in telecom, efforts remain severely limited. The most notable initiative is srsRANBench \cite{gajjar2025oransight}; however, it strictly measures static code understanding and repository comprehension, completely lacking any assessment of end-to-end code generation. To address domain-specific telecom challenges, several fine-tuned, telecom-adapted models have recently been developed, such as TelecomGPT \cite{zou2025telecomgpt} and ORANSight \cite{gajjar2025oransight}. Notably, ORANSight attempts to bridge the software gap by explicitly including the srsRAN codebase in its pretraining corpus. Yet, despite this domain-specific exposure, the practical application and evaluation of these systems remain restricted to code comprehension or the generation of isolated, single-function snippets. Current telecom AI frameworks do \textbf{not} assess the critical capabilities required for autonomous network management: end-to-end code generation, active multi-file codebase modification, cross-layer compilation, and the management of cascading execution logic necessary to maintain a softwarized protocol stack. TeleSWEBench directly addresses this void by moving beyond MCQA and static comprehension, providing the first rigorous testbed for functional, repository-scale code generation in telecommunications.

\section{TeleSWEBench}
\begin{figure*}[th]
    \centering
    \includegraphics[width=\linewidth]{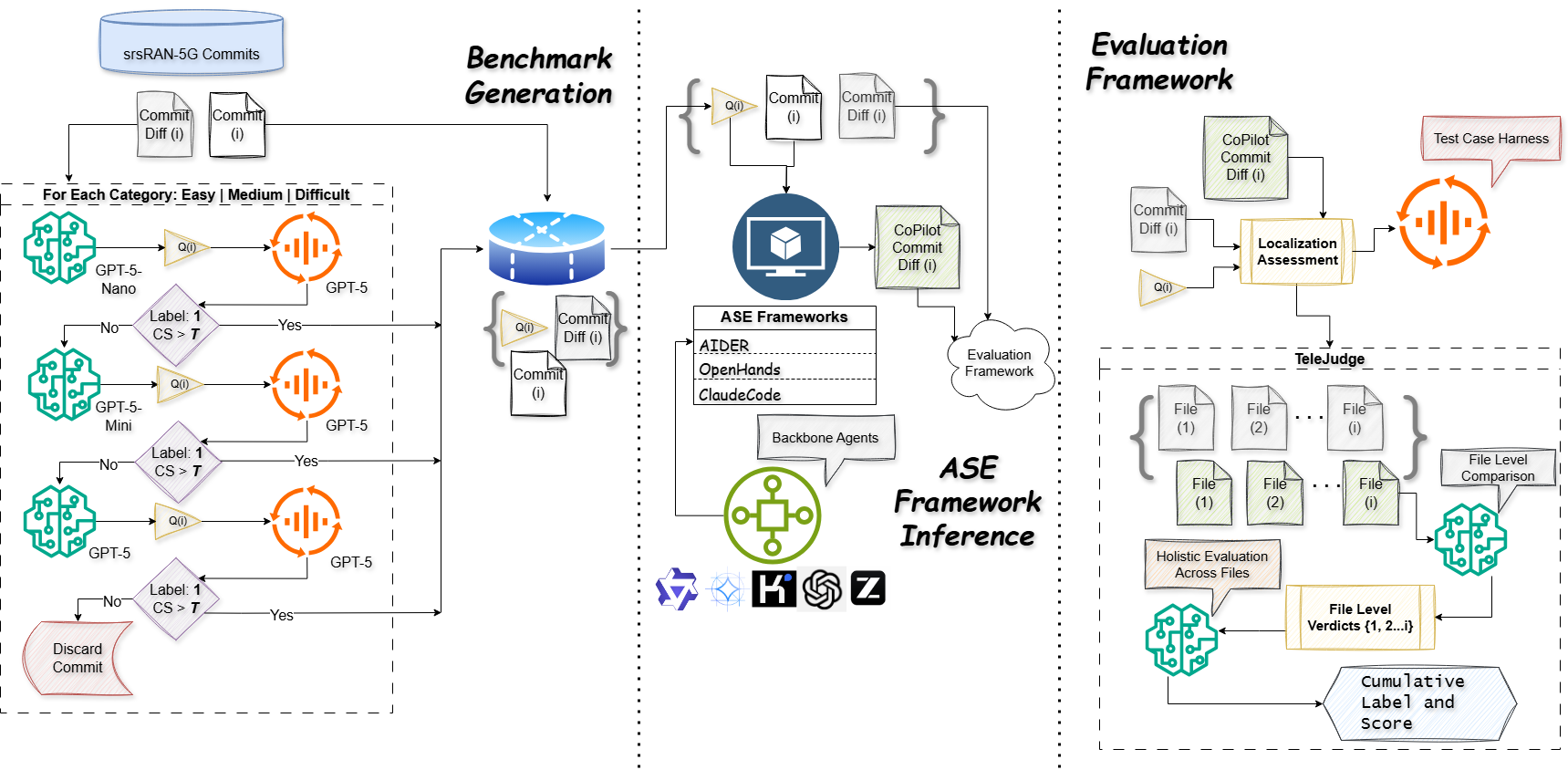}
    \caption{High-level overview of the TeleSWEBench from benchmark creation to evaluation.}
    \label{fig:overview}
\end{figure*}
We leverage \texttt{srsRAN 5G} as the foundation for our benchmark due to its prominence as an industry-standard open-source wireless stack. We sourced a comprehensive history of repository modifications spanning from 2023 to 2025, yielding a total of over 15k commits. Within this extensive history, the average number of file changes per commit stands at 5.57. This high degree of multi-file modification underscores the immense complexity and tightly coupled nature of telecommunication software. Unlike standard application development, where logic might be isolated to a single localized script, modifying a modern cellular protocol stack inherently requires synchronized updates across various interconnected layers, state machines, and hardware interfaces. Consequently, this statistical reality clearly demonstrates that evaluating autonomous software engineering agents in the telecommunications domain demands a benchmark capable of rigorously testing broad codebase navigation and deep repository context. By anchoring our evaluation on authentic developer activity with such high cross-file dependencies, we ensure that models are tested on the genuine architectural rigors they would face in real-world network deployment. A comprehensive overview of the TeleSWEBench pipeline, from commit mining to evaluation, is illustrated in Figure \ref{fig:overview}.

\subsection{Benchmark Generation}\label{benchg}
We partition benchmark items into three difficulty levels, each corresponding to how much of the solution is spelled out in the natural language task and how much the model must infer from the repository. \textit{Easy} tasks are written as near mechanical edit instructions. A valid easy prompt must give exact file paths and line references, specify precisely which code to remove and which code to insert, and remain fully self-contained so that, in principle, the change could be carried out with straightforward search and replace style operations without additional exploration. \textit{Medium} tasks sit between specification and discovery. They explain the \textit{what} and \textit{why} of the change, name the affected files or subsystems, and supply salient facts such as constants, function names, or version identifiers, but deliberately omit exact line-level edit recipes. Answering them is assumed to require some structural understanding of where the change belongs. \textit{Difficult} tasks state only high-level objectives and minimal hints such as a target value, endpoint, or version. They must not name specific file paths or functions, so the assistant must locate the right implementation context and plan edits autonomously. This tier is intended to stress repository scale reasoning rather than literal transcription of a patch description.

Generation is not fixed to a single frontier model. For each commit and each difficulty, we use a fallback chain over model depth, attempting generation once per model from smaller to larger before escalating. If a cheaper model produces a question that passes validation, we accept it and stop. Only on failure do we invoke the next, more capable and typically more expensive model. This \textit{cascade}, inspired by the paper \cite{jung2025trust}, keeps average generation cost low while still allowing difficult commits or strict validator outcomes to trigger stronger models when needed. For Validation, a language model is given the same commit context, the generated question, and the declared difficulty. The validator checks whether the text is unambiguous, carries an information density appropriate to that difficulty, is plausible as a task for an IDE coding assistant, is faithful to the actual commit diff, and is free of internal contradictions. In the confidence scoring variant, the validator additionally returns a calibrated style confidence score. We only retain items that are marked valid and meet a fixed \textit{confidence threshold} of $T$, which is kept as 0.9, which further reduces false positives in the benchmark. We apply a second heuristic on top of the model cascade. We generate and validate difficulties in the order \textit{difficult}, then medium, then easy, with early stopping. We only attempt medium if hard has already succeeded for that commit, and we only attempt easy if medium has succeeded. If \textit{difficult} fails, we abandon that commit for the benchmark. If hard succeeds but medium fails, we do not proceed to easy. 

This ordering reflects how constraining each tier is for both the generator and the validator. \textit{Difficult} questions allow vague, goal-oriented wording and forbid leaking file and function names, so it is intuitive to believe that they are relatively easy to phrase in a way that remains consistent with the commit. \textit{Easy} questions, by contrast, must satisfy the strongest structural requirements such as exact locations, exact removals and insertions, and full self containment, which are harder to generate correctly and harder to validate as faithful to the patch. \textit{Medium} sits between the two. By front-loading hard, we quickly filter commits that cannot support a coherent, difficult task. Furthermore, detailed prompts used for the LLM inference are available in the Appendix \ref{appx:prompts}

\subsection{ASE Framework Inference}

We specifically leverage Command Line Interface (CLI) based ASE frameworks, namely AIDER \cite{paul2026Aider}, OpenHands \cite{wang2025openhands}, and ClaudeCode \cite{liu2026dive}, with the former two being open source. These tools have rapidly established themselves as the standard multi-agent frameworks across numerous Automated Software Engineering (ASE) studies \cite{shen2025secrepobench, zhang2024diversity, wadhwa2024masai}, providing the essential scaffolding required for iterative repository navigation, contextual file reading, and autonomous patching. To ensure a fair and computationally feasible evaluation pipeline, we enforce a strict time-bound execution limit, defining a computational time cutoff $Time_m$ for all inference runs; if an agent fails to generate a final patch within this $Time_m$ threshold, the task is forcefully terminated and recorded as a failure. Powered by these frameworks, we benchmark a diverse suite of open weight backbone LLMs with SOTA performance in code generation and reasoning, across different parameter scales, including Qwen3.5, Gemma4, Kimi, GPT-OSS, and Qwencoder2.5, further explained in Section \ref{setup}. We evaluate backbone models from 1.5B to 1T parameters and varied context windows
to systematically identify how parameter scaling and extended context utilization impact an agent's success rate.

\subsection{Evaluation Framework}\label{sec:evaluation_framework}

We propose a two-stage evaluation pipeline, as shown in Figure \ref{fig:overview}, designed to explicitly decouple an agent's architectural navigation skills, identifying \textit{where} to make changes, from its ability to synthesize functionally correct domain logic. The first stage evaluates task localization, measuring whether the agent correctly understood the structural scope of the required modifications. Let $T$ represent the set of ground truth target files modified in the original developer commit, and $P$ represent the set of files modified by the ASE framework. We classify localization performance into one of five mutually exclusive states. An \textbf{Exact Match} occurs when the agent successfully isolates the exact required files ($P = T$). A \textbf{Partial Match} ($P \subset T$) indicates that the agent grasped elements of the problem and modified some correct files but failed to trace the required logic across the broader codebase. Conversely, an \textbf{Over Addressed} state ($T \subset P$) emerges when the agent correctly alters the target files but also inappropriately modifies unrelated files. Finally, we categorize complete localization failures as either a \textbf{No Match} ($P \cap T = \emptyset$, given $P \neq \emptyset$) when completely disjoint files are altered, or \textbf{No Changes} ($P = \emptyset$) when the agent merely generates text without any code changes.

The second stage evaluates functional correctness and is strictly reserved for outputs that achieve an Exact Match, ensuring that our assessment of code generation capabilities is not confounded by poor file localization. For these candidates, we first evaluate using a \textbf{Test Case Harness} across our benchmark, which consists of 734 questions that are accompanied by executable unit tests. We achieve this by natively compiling the agent's generated patch and executing it against the rigorous \texttt{srsRAN 5G} unit and integration test suites. However, rigid unit tests can often penalize the creativity of an SWE Agent, as they may fail functionally valid solutions that deviate structurally from the developer's original approach. To avoid penalizing this creativity, we further propose a LLM-as-a-Judge framework called \textbf{TeleJudge} that initially evaluates the diffs at the granularity of individual files. It then aggregates these micro-assessments into a holistic patch verdict (Pass or Fail) by reasoning about cross-file compatibility and structural coherence. We also evaluate other LLM-as-a-judge methods and show how these approaches often succumb to context overload and fail to produce reasonable output. Preliminary results regarding TeleJudge are available in Appendix \ref{appx:judge}. 


\section{Benchmark Statistics}

From the entire commit history, we can fetch all available commits that have repository-native unit tests, which comprise a small fraction of the entire commit database\footnote{More information regarding the test cases is available in the Appendix \ref{appx:tc}}. As the question generation process follows a heuristic, we can successfully generate Difficult questions for 313 commits, Medium questions for 279 commits, and Easy questions for 142 commits. The average number of files per question varies across these difficulty levels. Specifically, the \textit{Easy} split has a mean of 29.82 files and a median of 7. The Medium split features a mean of 19.57 files with a median of 4, while the Difficult split averages 18.42 files with a median of 5. Across all generated questions, the scope ranges from a minimum of 1 file to a maximum of 300 files per question. Sample questions along with the correct code changes are available in Appendix \ref{appx:samples_q}

\section{Experimental Setup}\label{setup}

All experiments were conducted on a dedicated workstation equipped with an Intel(R) Core(TM) i9-14900KF CPU, 62 GiB of RAM, and an NVIDIA GeForce RTX 4090 GPU with 24 GB of GDDR6X memory. We utilized the backbone models detailed in Table~\ref{tab:context_lengths}, employing the specific quantization formats available through the NAUTILUS framework \cite{nrp} to optimize for our local hardware limits. The only exception is the QwenCoder2.5, which is evaluated locally using Ollama \cite{Yang2026ollama}. Furthermore, to adhere to the constraints of our available compute resources, we enforced a strict maximum execution limit $Time_m$ of 300 seconds for the system to report a final answer. To first understand why the LLM Agents fail, we primarily leverage the \textbf{AIDER} framework because it is fully open-source and the most computationally efficient when compared to other open source options, including OpenHands \cite{ni2026gittaskbench}. Upon identifying the best-performing backbone model, we conducted an extended ablation study by replacing AIDER with \textbf{OpenHands} and \textbf{ClaudeCode}. This secondary evaluation phase allowed us to determine the extent to which the ASE framework and backbone architecture dictate telecom-specific success.
\begin{table}[ht]
\centering
\resizebox{\linewidth}{!}{%

\begin{tabular}{lccc}
\toprule
\textbf{Model} & \textbf{Parameters} & \textbf{CL} & \textbf{Quant} \\
\midrule

Qwen 3.5 \cite{qwen35blog} & 397B & 1M & FP8 \\
Qwen 3.5 Small \cite{qwen35blog} & 35B & 1M & FP8 \\
GPT-OSS \cite{openai2025gptoss120bgptoss20bmodel} & 120B & 131K & MXFP4 \\
Gemma 4 \cite{farabet_2026_gemma} & 31B & 262K & FP16 \\
Kimi K2.5 \cite{kimiteam2026kimik25visualagentic} & 1T & 262K & MXFP4 \\
QwenCoder 2.5 \cite{hui2024qwen25codertechnicalreport} & 1.5B & 32K & Int4 \\
\bottomrule
\end{tabular}}
\caption{Parameters, maximum context lengths (CL), and quantization formats (Quant) for the evaluated backbone LLMs.}
\label{tab:context_lengths}
\end{table}

\textbf{Metrics} Building upon the evaluation framework established in Section \ref{sec:evaluation_framework}, we formalize our assessment into distinct metrics for both stages of the pipeline. For Stage 1, we primarily measure task localization using the \textit{Exact Match} (EM) rate, which represents the proportion of questions where the correct files are modified ($P = T$). Given $N$ total questions and $N_{\text{EM}}$ instances of exact matches, EM is defined as $\text{EM} = N_{\text{EM}} / N$. Similarly, we calculate the rates for Partial Match (PM), Over Addressed (OA), No Match (NM), and No Changes (NC) by dividing their respective occurrence counts by the total number of questions $N$. For Stage 2, we evaluate the code generation capabilities strictly on the subset of $N_{\text{EM}}$ candidates. To measure functional correctness, we define the \textit{Unit Test Acceptance Rate} (UAR). If $N_{\text{UT}}$ represents the number of generated patches that successfully compile and pass all associated unit and integration tests, UAR is calculated as $\text{UAR} = N_{\text{UT}} / N_{\text{EM}}$. Finally, to quantify the outputs processed by our TeleJudge framework, we define the \textit{TeleJudge Acceptance Rate} (TAR). This metric identifies the proportion of structurally creative and accurate solutions that pass the holistic verdict. Letting $N_{\text{pass}}$ denote the number of patches accepted by the judge for file level and finally the holistic verdict, TAR is given by $\text{TAR} = N_{\text{pass}} / N_{\text{EM}}$. To capture end-to-end deployability, we introduce the \textit{Ship-Ready Percentage} (SRP), which jointly requires semantic acceptance and executable correctness on the same exact-match subset. Let $N_{\text{SR}}$ denote the number of Stage~2 candidates (from $N_{\text{EM}}$) whose patches are \emph{accepted} by TeleJudge and also \emph{pass} all required build and test checks. Then $\text{SRP} \;=\; \frac{N_{\text{SR}}}{N_{\text{EM}}}$.
Formally, let $i \in \{1,\dots,N_{\text{EM}}\}$ index the exact-match instances, which gives us $N_{\text{SR}} \;=\; \sum_{i=1}^{N_{\text{EM}}} \mathbf{1}\!\left[\text{JudgeAccept}_i \wedge \text{TestPass}_i\right]$ so SRP measures the proportion of candidates that are simultaneously holistically correct and functionally verified. This makes SRP a stricter end-to-end metric than either UAR or TAR alone.
\begin{table*}[h]
    \centering
    \caption{Stage 1 Localization Results across Difficulty Tiers. Runs that ended in \texttt{timeout} are excluded.}
    \label{tab:stage1_localization}
    \resizebox{\textwidth}{!}{%
    \begin{tabular}{l | ccccc | ccccc | ccccc | ccccc}
        \toprule
        \multirow{2}{*}{\textbf{Model}} & \multicolumn{5}{c|}{\textbf{Easy}} & \multicolumn{5}{c|}{\textbf{Medium}} & \multicolumn{5}{c|}{\textbf{Difficult}} & \multicolumn{5}{c}{\textbf{Cumulative}} \\
        & EM & PM & OA & NM & NC & EM & PM & OA & NM & NC & EM & PM & OA & NM & NC & EM & PM & OA & NM & NC \\
        \midrule
        \texttt{Qwen3}         & 24.0 & 19.2 & 0.0 & 0.0 & 56.7 & 25.0 & 27.0 & 3.2 & 4.8 & 39.9 & 1.3 & 6.7 & 0.7 & 16.2 & 75.1 & 14.0 & 16.5 & 1.5 & 9.2 & 58.7 \\
        \texttt{Qwen3-small}         & 1.6 & 11.9 & 0.8 & 0.8 & 84.9 & 7.4 & 14.8 & 0.0 & 2.6 & 75.3 & 1.0 & 2.3 & 0.0 & 8.7 & 88.0 & 3.5 & 8.8 & 0.1 & 5.0 & 82.6 \\
        \texttt{Kimi-K2.5}         & 5.3 & 22.8 & 0.0 & 0.0 & 71.9 & 13.8 & 19.5 & 0.8 & 4.1 & 61.8 & 0.9 & 6.4 & 0.0 & 6.0 & 86.8 & 5.3 & 12.6 & 0.2 & 4.6 & 77.3 \\
        \texttt{Gemma4}         & 1.5 & 9.2 & 0.0 & 0.0 & 89.2 & 7.6 & 16.5 & 0.0 & 0.6 & 75.3 & 0.3 & 1.4 & 0.0 & 5.4 & 92.9 & 2.8 & 7.2 & 0.0 & 3.2 & 86.8 \\
        \texttt{GPT-OSS}         & 19.7 & 13.1 & 0.8 & 1.6 & 64.8 & 19.9 & 24.0 & 2.6 & 4.5 & 49.1 & 0.0 & 5.4 & 0.0 & 14.4 & 80.2 & 11.0 & 13.8 & 1.1 & 8.4 & 65.7 \\
        \texttt{QwenCoder-2.5}         & 37.8 & 21.9 & 2.1 & 20.3 & 17.9 & 20.4 & 6.4 & 1.9 & 32.6 & 38.7 & 5.3 & 4.0 & 1.7 & 43.5 & 45.5 & 17.4 & 8.1 & 1.8 & 34.8 & 37.8 \\        \texttt{GLM-4.7}         & 3.2 & 29.0 & 0.0 & 0.0 & 67.7 & 6.0 & 4.0 & 0.0 & 0.0 & 90.0 & 0.0 & 2.6 & 0.0 & 0.5 & 96.8 & 1.5 & 5.9 & 0.0 & 0.4 & 92.2 \\
        \bottomrule
    \end{tabular}%
    }
\end{table*}

\section{Results and Analysis}

\subsection{Stage 1 Localization Performance}\label{sec:stage1}

Table \ref{tab:stage1_localization} presents the comprehensive results for the initial localization stage across all difficulty tiers. A clear trend emerges regarding the inverse relationship between task complexity and localization success. Across all models evaluated, the Exact Match (EM) rates degrade significantly as the scope of the tasks shifts from Easy to Difficult. Concurrently, the No Match (NM) and No Changes (NC) metrics exhibit a sharp increase in the more complex tiers. For example, QwenCoder 2.5 achieves a 37.8\% EM rate on Easy tasks, but this performance drops to a mere 5.3\% on Difficult tasks, accompanied by a corresponding surge in its NM rate to 43.5\%. This demonstrates that as the necessary context spans deeper across the repository hierarchy without explicit file hints, models increasingly fail to isolate the target logic. Interestingly, our evaluation highlights a counterintuitive dynamic regarding model scale and localization efficacy. The smallest model in our evaluation suite actually localizes remarkably well and acts more decisively to modify the codebase. In contrast, the larger, structurally more capable models demonstrate a highly reserved performance profile. This is heavily reflected in their disproportionate NC rates; for instance, GLM-4.7 and Gemma4 output cumulative NC rates of 92.2\% and 86.8\%, respectively. We hypothesize that this phenomenon is due to the larger foundation models being overly aligned for safety and conversational timidity. When faced with the dense, interconnected C++ architecture of the telecom repository, these models become overly cautious and refuse to execute concrete modifications. We provide qualitative examples of these interactions in the Appendix \ref{appx:timmid}.

\subsubsection{Failure Points}

\begin{figure}[h]
    \centering
    \includegraphics[width=\linewidth]{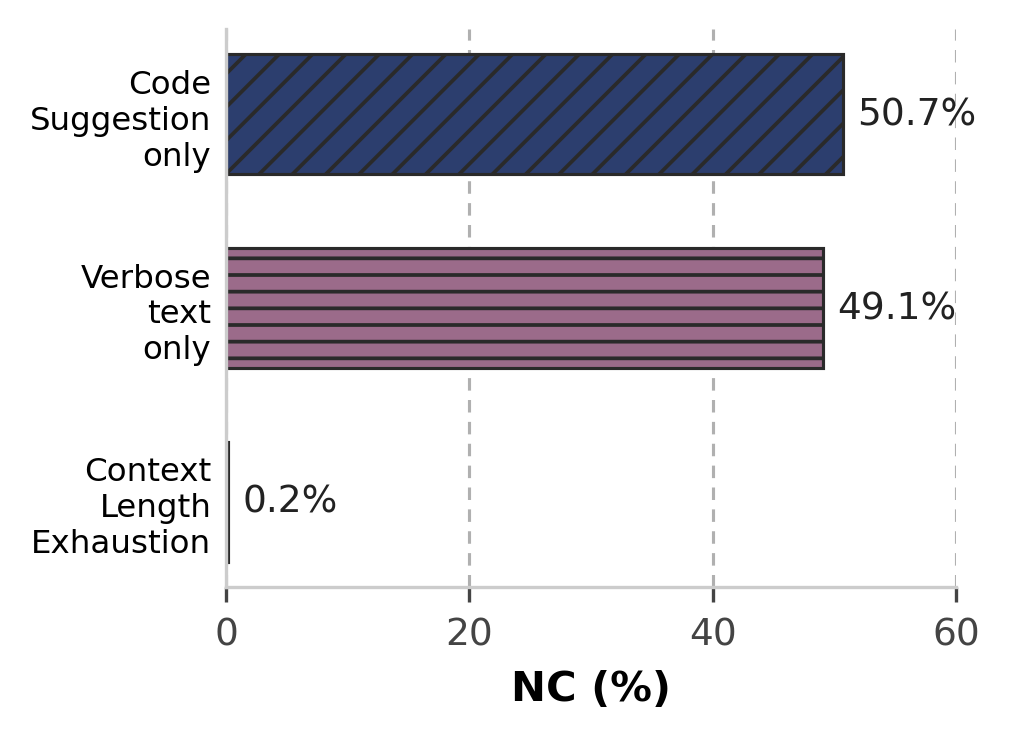}
    \caption{Distribution for the three failure points across all \textit{NC} outputs.}
    \label{fig:failure}
\end{figure}

Beyond the quantitative performance drops observed in the localization metrics, we identified three distinct behavioral failure modes that prevent models from successfully resolving tasks. As illustrated in Figure \ref{fig:failure}, the first prevalent point of failure across the \textit{modern} LLMs (except \texttt{QwenCoder-2.5}) occurs when the model successfully deduces which files and lines require modification, but strictly outputs plain-text suggestions or pseudocode rather than generating the actionable code edits, resulting in a \textit{timid} behavior. The second failure mode involves models generating excessively verbose outputs that lack actionable code suggestions, frequently causing the model to halt execution to ask clarifying counter-questions instead of proceeding autonomously. Finally, the third failure point is context exhaustion. Given the massive scale of the srsRAN 5G repository, the iterative file exploration required for Medium and Difficult tasks frequently overwhelms the context windows of the agents, leading to truncated reasoning loops and aborted runs before any edits can be localized.

\subsection{Stage 2 Functional Correctness}

\begin{figure}[t]
    \centering
    \includegraphics[width=\linewidth]{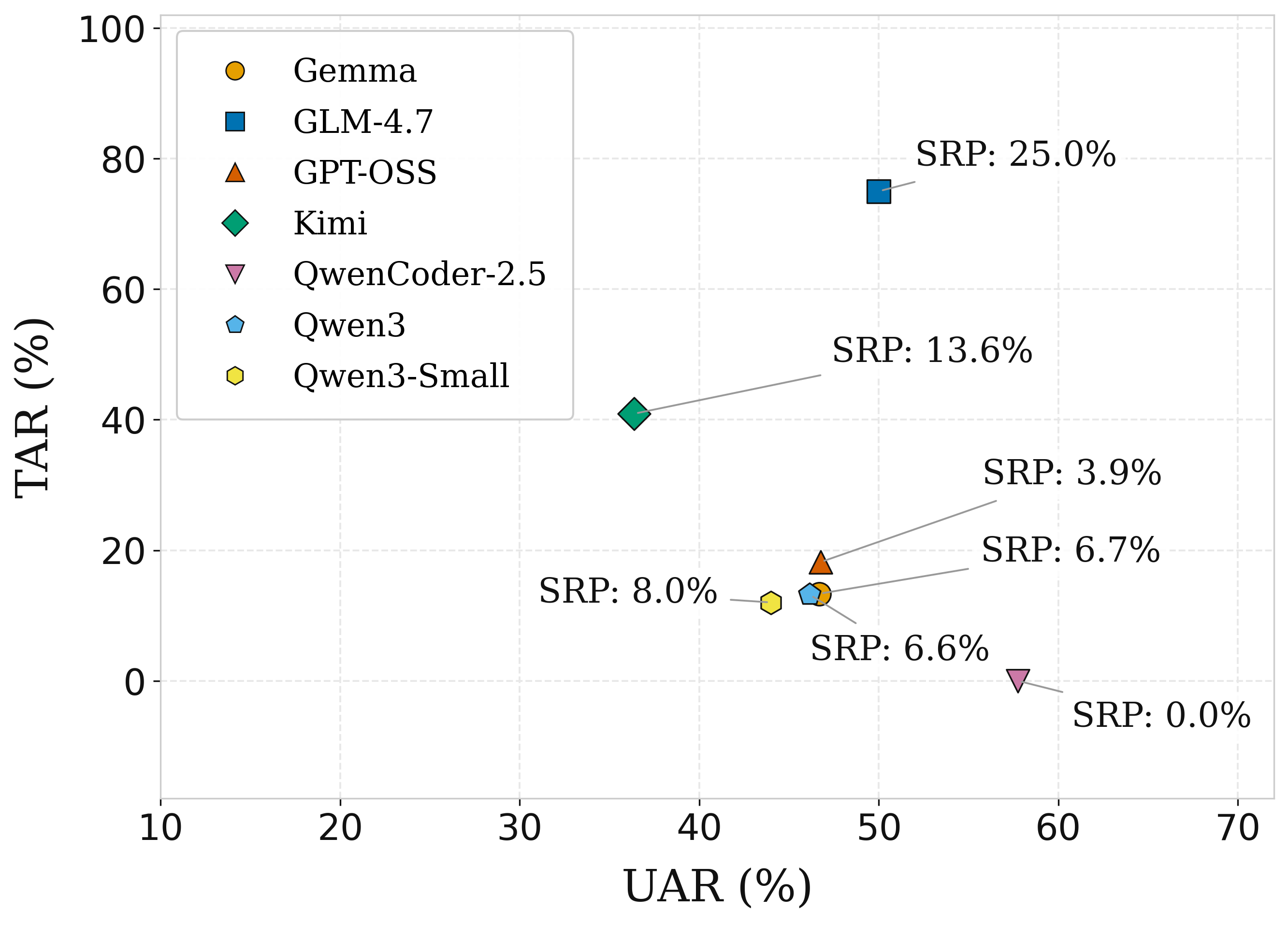}
    \caption{Each marker represents a model plotted by \textit{UAR} on the x axis and \textit{TAR} on the y axis and the annotations report the \textit{SRP}.}
    \label{fig:spr}
\end{figure}

By restricting our Stage 2 evaluation strictly to instances where models achieved an Exact Match in localization, we assess true functional correctness through three complementary metrics: UAR, TAR, and SRP. Relying solely on unit tests or qualitative judgments fails to capture true deployment readiness, making this joint evaluation critical for assessing telecom code quality. Figure \ref{fig:spr} illustrates a pronounced gap between executable and semantic correctness across the evaluated models.
Crucially, it must be noted that the overall absolute performance across all models remains exceedingly low; our analysis here primarily explores relative behavioral differences rather than identifying a genuinely deployment-ready system. \texttt{GLM-4.7} emerges as the strongest relative performer, yet it only achieves an end-to-end SRP of 25.0\%. However, this result directly contextualizes the timidity barrier identified in our Stage 1 analysis. Although \texttt{GLM-4.7} exhibited a massive NC rate exceeding 90\%, this extreme caution translates into higher relative precision. On the rare occasions when the model actually commits to an edit and correctly localizes the scope, its solutions are comparatively reliable and holistically sound. In contrast, \texttt{QwenCoder-2.5} embodies the exact opposite behavioral profile. While Stage 1 showed it to be highly decisive and willing to modify code, its Stage 2 performance reveals severe operational brittleness. The model achieves a relatively strong UAR (57.73\%), successfully compiling code that passes the native test suites, but it completely fails the TeleJudge semantic evaluation with a TAR of 0.0\%, resulting in an SRP of 0.0\%. This discrepancy indicates that the model frequently generates narrow workarounds that blindly satisfy isolated unit tests but catastrophically violate broader cross-file compatibility, stateful logic, and strict C++ telecom conventions. The remaining models fall between these two extremes, none of which perform exceptionally well. \texttt{Kimi-K2.5} demonstrates moderate alignment across all metrics (\(\mathrm{UAR}=36.36\%\), \(\mathrm{TAR}=40.91\%\), \(\mathrm{SRP}=13.64\%\)). Meanwhile, models like \texttt{Qwen3-small} and \texttt{Gemma} exhibit comparable test execution behavior with \(\mathrm{UAR}\approx 44\!-\!47\%\) but suffer from drastically lower judge acceptance, yielding single-digit SRP scores. Ultimately, the persistent divergence between TAR and UAR underscores that passing unit tests alone is vastly insufficient for automated software engineering in the telecom domain. 

\subsection{OpenHands and ClaudeCode}

\begin{figure}
    \centering
    \includegraphics[width=\linewidth]{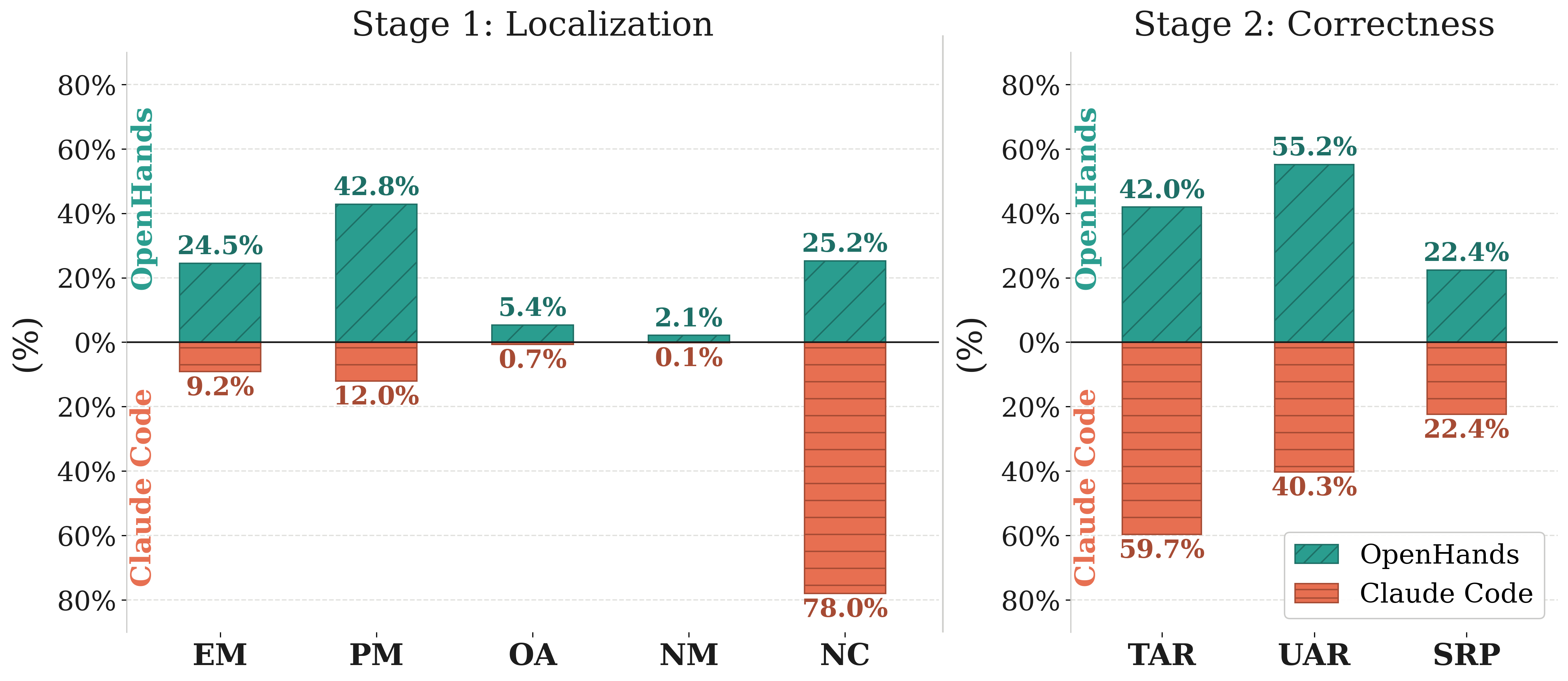}
    \caption{Performance comparison between OpenHands and ClaudeCode ASE frameworks using the \texttt{qwen3} backbone.}
    \label{fig:openhands_comparison}
\end{figure}

To further investigate the impact of an ASE framework on code generation within the telecommunications domain, we conducted an extended case study utilizing the OpenHands and ClaudeCode frameworks backed by our top-performing foundation model from Table \ref{tab:stage1_localization}, \texttt{qwen3}. We do observe that \texttt{QwenCoder-2.5} provides a better EM rate, but due to a 0\% $\mathrm{SRP}$, we choose \texttt{qwen3}. As shown in Figure \ref{fig:openhands_comparison}, the performance varies significantly across agent architectures. Utilizing AIDER resulted in a TAR of 13.2, a UAR of 46.2, and an overall $\mathrm{SRP}$ of 6.6. The alternatives delivered superior results, confirming that the ASE framework heavily influences efficacy. Additionally, AIDER recorded an EM rate of 14.0 and an NC rate of 58.7. These eclipse ClaudeCode, a result we attribute to \textit{timidness}, where the tool frequently avoids making changes but maintains high correctness when it does modify the codebase.

\section{Conclusion}
TeleSWEBench bridges a critical gap in automated software engineering evaluation by introducing the first quantitative commit-driven benchmark tailored specifically for the telecommunications domain. By mining real-world developer commits, we constructed 734 structured test cases categorized into three distinct difficulty tiers to rigorously evaluate agent capabilities. Our comprehensive two-stage evaluation assessing both task localization and functional correctness reveals that current agents severely lack the domain-specific reasoning required to reliably patch complex wireless stacks like \texttt{srsRAN 5G}. Even the most advanced ASE frameworks struggled significantly, with the best-performing tools achieving up to 25\% success rate for producing shippable code. We observed that while models can sometimes handle basic tasks, their performance degrades drastically on difficult tiers where they frequently fail due to context exhaustion, timidity, or a tendency to output plain text explanations instead of executable code edits. 

For future work, we wish to address the Limitations mentioned below by expanding the benchmark to a wider array of repositories and broadening our baselines to encompass commercial LLM models and ASE frameworks. 

\section*{Limitations}\label{limitations}

\textbf{Single Repository}. Currently, our benchmark is built exclusively around the srsRAN 5G repository. While srsRAN serves as a highly robust, representative, and widely used open-source telecom stack, restricting our dataset to a single codebase may narrow the scope of our findings. The unique architectural choices, coding conventions, and documentation styles specific to srsRAN might not fully capture the diverse complexities present in other open-source telecom projects or proprietary, industry-grade software, potentially limiting the broader generalizability of the evaluated agents.

\textbf{Evaluated Frameworks}. Furthermore, our baseline evaluations are restricted to only three ASE frameworks, omitting other closed-source or commercial alternatives. This decision was strictly necessitated by the prohibitive costs associated with executing large-scale, iterative software engineering tasks, which often require continuous API calls, environment interactions, and extensive token generation on paid, proprietary models. Although including state-of-the-art closed-source frameworks might yield higher performance baselines, utilizing open-source frameworks ensures that our current evaluation pipeline remains fully reproducible and financially accessible to the broader research community.

\textbf{Reliance on Existing Test Cases}. The unit tests utilized in our benchmark are extracted directly from the \texttt{srsRAN-5G} repository rather than being custom-authored specifically for each evaluation task. Because we rely on these pre-existing developer tests, they may not be perfectly tailored to catch every edge case or unintended side effect introduced by an AI agent's specific modifications. As a result, the functional correctness evaluation might occasionally pass flawed code or fail structurally unconventional but functional code.

\textbf{Infrastructure Constraints and Timeouts}. Finally, as detailed in Section \ref{setup}, our experiments relied on the NAUTILUS platform for model API access. Because NAUTILUS does not provide strict Service Level Agreements (SLAs), API calls occasionally drop or exit unexpectedly, leading to premature ASE tool timeouts. While hosting these large models locally could have mitigated these network-related failures and potentially allowed agents to fully complete their coding tasks, our limited compute resources precluded large-scale local deployment. Consequently, our Stage 2 performance evaluation is only conducted on the Exact Match (EM) samples where generation was completed, and the appripriate files were changed.

\bibliography{refs}

\begin{thebibliography}{40}
\providecommand{\natexlab}[1]{#1}

\bibitem[{Austin et~al.(2021)Austin, Odena, Nye, Bosma, Michalewski, Dohan, Jiang, Cai, Terry, Le et~al.}]{austin2021program}
Jacob Austin, Augustus Odena, Maxwell Nye, Maarten Bosma, Henryk Michalewski, David Dohan, Ellen Jiang, Carrie Cai, Michael Terry, Quoc Le, and 1 others. 2021.
\newblock Program synthesis with large language models.
\newblock \emph{arXiv preprint arXiv:2108.07732}.

\bibitem[{Baron and Gupta(2018)}]{baron2018unpacking}
Justus Baron and Kirti Gupta. 2018.
\newblock Unpacking 3gpp standards.
\newblock \emph{Journal of Economics \& Management Strategy}, 27(3):433--461.

\bibitem[{Chen et~al.(2021)Chen, Tworek, Jun, Yuan, Pinto, Kaplan, Edwards, Burda, Joseph, Brockman et~al.}]{chen2021evaluating}
Mark Chen, Jerry Tworek, Heewoo Jun, Qiming Yuan, Henrique Ponde De~Oliveira Pinto, Jared Kaplan, Harri Edwards, Yuri Burda, Nicholas Joseph, Greg Brockman, and 1 others. 2021.
\newblock Evaluating large language models trained on code.
\newblock \emph{arXiv preprint arXiv:2107.03374}.

\bibitem[{Chowdhury et~al.(2024)Chowdhury, Aung, Shern, Jaffe, Sherburn, Starace, Mays, Dias, Aljubeh, Glaese, Jimenez, Yang, Ho, Patwardhan, Liu, and Madry}]{chowdhury2024swebenchverified}
Neil Chowdhury, James Aung, Chan~Jun Shern, Oliver Jaffe, Dane Sherburn, Giulio Starace, Evan Mays, Rachel Dias, Marwan Aljubeh, Mia Glaese, Carlos~E. Jimenez, John Yang, Leyton Ho, Tejal Patwardhan, Kevin Liu, and Aleksander Madry. 2024.
\newblock \href {https://openai.com/index/introducing-swe-bench-verified/} {Introducing {SWE}-bench verified}.

\bibitem[{Farabet(2026)}]{farabet_2026_gemma}
Clement Farabet. 2026.
\newblock \href {https://blog.google/innovation-and-ai/technology/developers-tools/gemma-4/} {Gemma 4: Byte for byte, the most capable open models}.

\bibitem[{Feng et~al.(2026)Feng, Yang, Guo, Xia, Liu, and Quek}]{FENG20265}
Chenyuan Feng, Howard~H. Yang, Kun Guo, Wenchao Xia, Chenxi Liu, and Tony~Q.S. Quek. 2026.
\newblock \href {https://doi.org/10.1016/j.jiixd.2025.12.013} {Ai-ran: The pathway to future wireless networks}.
\newblock \emph{Journal of Information and Intelligence}, 4(1):5--22.

\bibitem[{Gajjar and Shah(2025{\natexlab{a}})}]{gajjar2025oran}
Pranshav Gajjar and Vijay~K Shah. 2025{\natexlab{a}}.
\newblock Oran-bench-13k: An open source benchmark for assessing llms in open radio access networks.
\newblock In \emph{2025 IEEE 22nd Consumer Communications \& Networking Conference (CCNC)}, pages 1--4. IEEE.

\bibitem[{Gajjar and Shah(2025{\natexlab{b}})}]{gajjar2025oransight}
Pranshav Gajjar and Vijay~K Shah. 2025{\natexlab{b}}.
\newblock Oransight-2.0: Foundational llms for o-ran.
\newblock \emph{IEEE Transactions on Machine Learning in Communications and Networking}.

\bibitem[{Ganiyu et~al.(2025)Ganiyu, Gajjar, and Shah}]{ganiyu2025ai5gtest}
Abiodun Ganiyu, Pranshav Gajjar, and Vijay~K Shah. 2025.
\newblock Ai5gtest: Ai-driven specification-aware automated testing and validation of 5g o-ran components.
\newblock In \emph{18th ACM Conference on Security and Privacy in Wireless and Mobile Networks}, pages 53--64.

\bibitem[{GSMA(2026)}]{gsma_release_2026}
GSMA. 2026.
\newblock \href {https://github.com/gsma-labs/evals/releases/tag/v1.1.0} {Release v1.1.0 · gsma-labs/evals}.

\bibitem[{Gu et~al.(2024)Gu, Jiang, Shi, Tan, Zhai, Xu, Li, Shen, Ma, Liu et~al.}]{gu2024survey}
Jiawei Gu, Xuhui Jiang, Zhichao Shi, Hexiang Tan, Xuehao Zhai, Chengjin Xu, Wei Li, Yinghan Shen, Shengjie Ma, Honghao Liu, and 1 others. 2024.
\newblock A survey on llm-as-a-judge.
\newblock \emph{The Innovation}.

\bibitem[{He et~al.(2026)He, Shi, Zhuo, Treude, Sun, Xing, Du, and Lo}]{he2026llm}
Junda He, Jieke Shi, Terry~Yue Zhuo, Christoph Treude, Jiamou Sun, Zhenchang Xing, Xiaoning Du, and David Lo. 2026.
\newblock Llm-as-a-judge for software engineering: Literature review, vision, and the road ahead.
\newblock \emph{ACM Transactions on Software Engineering and Methodology}.

\bibitem[{Hou et~al.(2024)Hou, Zhao, Liu, Yang, Wang, Li, Luo, Lo, Grundy, and Wang}]{hou2024large}
Xinyi Hou, Yanjie Zhao, Yue Liu, Zhou Yang, Kailong Wang, Li~Li, Xiapu Luo, David Lo, John Grundy, and Haoyu Wang. 2024.
\newblock Large language models for software engineering: A systematic literature review.
\newblock \emph{ACM Transactions on Software Engineering and Methodology}, 33(8):1--79.

\bibitem[{Hui et~al.(2024)Hui, Yang, Cui, Yang, Liu, Zhang, Liu, Zhang, Yu, Lu, Dang, Fan, Zhang, Yang, Men, Huang, Zheng, Miao, Quan, Feng, Ren, Ren, Zhou, and Lin}]{hui2024qwen25codertechnicalreport}
Binyuan Hui, Jian Yang, Zeyu Cui, Jiaxi Yang, Dayiheng Liu, Lei Zhang, Tianyu Liu, Jiajun Zhang, Bowen Yu, Keming Lu, Kai Dang, Yang Fan, Yichang Zhang, An~Yang, Rui Men, Fei Huang, Bo~Zheng, Yibo Miao, Shanghaoran Quan, and 5 others. 2024.
\newblock \href {https://arxiv.org/abs/2409.12186} {Qwen2.5-coder technical report}.
\newblock \emph{Preprint}, arXiv:2409.12186.

\bibitem[{Jimenez et~al.(2023)Jimenez, Yang, Wettig, Yao, Pei, Press, and Narasimhan}]{jimenez2023swe}
Carlos~E Jimenez, John Yang, Alexander Wettig, Shunyu Yao, Kexin Pei, Ofir Press, and Karthik Narasimhan. 2023.
\newblock Swe-bench: Can language models resolve real-world github issues?
\newblock \emph{arXiv preprint arXiv:2310.06770}.

\bibitem[{Jung et~al.(2025)Jung, Brahman, and Choi}]{jung2025trust}
Jaehun Jung, Faeze Brahman, and Yejin Choi. 2025.
\newblock \href {https://openreview.net/forum?id=UHPnqSTBPO} {Trust or escalate: {LLM} judges with provable guarantees for human agreement}.
\newblock In \emph{The Thirteenth International Conference on Learning Representations}.

\bibitem[{Li et~al.(2025)Li, Jiang, Huang, Beigi, Zhao, Tan, Bhattacharjee, Jiang, Chen, Wu et~al.}]{li2025generation}
Dawei Li, Bohan Jiang, Liangjie Huang, Alimohammad Beigi, Chengshuai Zhao, Zhen Tan, Amrita Bhattacharjee, Yuxuan Jiang, Canyu Chen, Tianhao Wu, and 1 others. 2025.
\newblock From generation to judgment: Opportunities and challenges of llm-as-a-judge.
\newblock In \emph{Proceedings of the 2025 Conference on Empirical Methods in Natural Language Processing}, pages 2757--2791.

\bibitem[{Li et~al.(2024)Li, Dong, Chen, Su, Zhou, Ai, Ye, and Liu}]{li2024llms}
Haitao Li, Qian Dong, Junjie Chen, Huixue Su, Yujia Zhou, Qingyao Ai, Ziyi Ye, and Yiqun Liu. 2024.
\newblock Llms-as-judges: a comprehensive survey on llm-based evaluation methods.
\newblock \emph{arXiv preprint arXiv:2412.05579}.

\bibitem[{Liu et~al.(2026)Liu, Zhao, Shang, and Shen}]{liu2026dive}
Jiacheng Liu, Xiaohan Zhao, Xinyi Shang, and Zhiqiang Shen. 2026.
\newblock Dive into claude code: The design space of today's and future ai agent systems.
\newblock \emph{arXiv preprint arXiv:2604.14228}.

\bibitem[{Liu et~al.(2023)Liu, Xu, and McAuley}]{liu2023repobench}
Tianyang Liu, Canwen Xu, and Julian McAuley. 2023.
\newblock Repobench: Benchmarking repository-level code auto-completion systems.
\newblock \emph{arXiv preprint arXiv:2306.03091}.

\bibitem[{Maatouk et~al.(2025)Maatouk, Ayed, Piovesan, De~Domenico, Debbah, and Luo}]{maatouk2025teleqna}
Ali Maatouk, Fadhel Ayed, Nicola Piovesan, Antonio De~Domenico, Merouane Debbah, and Zhi-Quan Luo. 2025.
\newblock Teleqna: A benchmark dataset to assess large language models telecommunications knowledge.
\newblock \emph{IEEE Network}.

\bibitem[{Ni et~al.(2026)Ni, Wang, Zhang, Lu, He, Tang, Hu, Li, Hu, Jiao et~al.}]{ni2026gittaskbench}
Ziyi Ni, Huacan Wang, Shuo Zhang, Shuo Lu, Ziyang He, Zhenheng Tang, Sen Hu, Bo~Li, Chen Hu, Binxing Jiao, and 1 others. 2026.
\newblock Gittaskbench: A benchmark for code agents solving real-world tasks through code repository leveraging.
\newblock In \emph{Proceedings of the AAAI Conference on Artificial Intelligence}, volume~40, pages 32564--32572.

\bibitem[{OpenAI et~al.(2025)OpenAI, :, Agarwal, Ahmad, Ai, Altman, Applebaum, Arbus, Arora, Bai, Baker, Bao, Barak, Bennett, Bertao, Brett, Brevdo, Brockman, Bubeck, Chang, Chen, Chen, Cheung, Clark, Cook, Dukhan, Dvorak, Fives, Fomenko, Garipov, Georgiev, Glaese, Gogineni, Goucher, Gross, Guzman, Hallman, Hehir, Heidecke, Helyar, Hu, Huet, Huh, Jain, Johnson, Koch, Kofman, Kundel, Kwon, Kyrylov, Le, Leclerc, Lennon, Lessans, Lezcano-Casado, Li, Li, Lin, Liss, Lily, Liu, Liu, Lu, Lu, Martinovic, McCallum, McGrath, McKinney, McLaughlin, Mei, Mostovoy, Mu, Myles, Neitz, Nichol, Pachocki, Paino, Palmie, Pantuliano, Parascandolo, Park, Pathak, Paz, Peran, Pimenov, Pokrass, Proehl, Qiu, Raila, Raso, Ren, Richardson, Robinson, Rotsted, Salman, Sanjeev, Schwarzer, Sculley, Sikchi, Simon, Singhal, Song, Stuckey, Sun, Tillet, Toizer, Tsimpourlas, Vyas, Wallace, Wang, Wang, Watkins, Weil, Wendling, Whinnery, Whitney, Wong, Yang, Yang, Yasunaga, Ying, Zaremba, Zhan, Zhang, Zhang, Zhang, and
  Zhao}]{openai2025gptoss120bgptoss20bmodel}
OpenAI, :, Sandhini Agarwal, Lama Ahmad, Jason Ai, Sam Altman, Andy Applebaum, Edwin Arbus, Rahul~K. Arora, Yu~Bai, Bowen Baker, Haiming Bao, Boaz Barak, Ally Bennett, Tyler Bertao, Nivedita Brett, Eugene Brevdo, Greg Brockman, Sebastien Bubeck, and 108 others. 2025.
\newblock \href {https://arxiv.org/abs/2508.10925} {gpt-oss-120b \& gpt-oss-20b model card}.
\newblock \emph{Preprint}, arXiv:2508.10925.

\bibitem[{Paisana et~al.(2026)Paisana, Eckermann, Alvarez, Falkenberg, Arteaga, S{\' a}ez, Herle, {carlo-gal}, {AlaiaL}, Puschmann, {dvdgrgrtt}, {faluco}, {jonathan-srs}, Broquetas, Tallon, {Piotr}, {Pavel}, Font-Bach, F{\" u}rste, {adrian}, McAuliffe, Sutton, Codebot, Carrera, {jcgonzalez-srs}, Apostolakis, {CedricRouxEurecom}, {ximbisrs}, and {qbroquetas}}]{Paisana2026srsran}
Francisco Paisana, Fabian Eckermann, Pedro Alvarez, Robert Falkenberg, Xavier Arteaga, Alfredo S{\' a}ez, Supreeth Herle, {carlo-gal}, {AlaiaL}, Andre Puschmann, {dvdgrgrtt}, {faluco}, {jonathan-srs}, Joaquim Broquetas, Justin Tallon, {Piotr}, {Pavel}, Oriol Font-Bach, Nils F{\" u}rste, and 10 others. 2026.
\newblock \href {https://github.com/srsran/srsRAN_Project} {srsran/{srsRAN}\textunderscore{}{Project}}.
\newblock https://github.com/srsran/srsRAN\textunderscore{}Project.

\bibitem[{{paul-gauthier} et~al.(2026){paul-gauthier}, Grigorev, Vial, {fry69}, {Krazer}, {shladnik}, Lackey, {Titusz}, Pellegrino, Vainsencher, {xqyz}, Schilling, Alammar, Kaihola, Landi, Ahmad, {MDW}, Dizi, Fioravanti, Cheah, {youknow}, {pcamp}, ALI, Shackelford, Dengler, Nguyen, Zhang, {therealmarv}, Muraviev, and {hypn4}}]{paul2026Aider}
{paul-gauthier}, Andrew Grigorev, Joshua Vial, {fry69}, {Krazer}, {shladnik}, IT~Lackey, {Titusz}, Claudia Pellegrino, Daniel Vainsencher, {xqyz}, Peter Schilling, Jay Alammar, Antti Kaihola, Matteo Landi, Farhan Ahmad, {MDW}, Kenny Dizi, Ivan Fioravanti, and 11 others. 2026.
\newblock \href {https://github.com/Aider-AI/aider} {Aider-{AI}/aider}.
\newblock https://github.com/Aider-AI/aider.

\bibitem[{Polese et~al.(2023)Polese, Bonati, D’oro, Basagni, and Melodia}]{polese2023understanding}
Michele Polese, Leonardo Bonati, Salvatore D’oro, Stefano Basagni, and Tommaso Melodia. 2023.
\newblock Understanding o-ran: Architecture, interfaces, algorithms, security, and research challenges.
\newblock \emph{IEEE Communications Surveys \& Tutorials}, 25(2):1376--1411.

\bibitem[{Shen et~al.(2025)Shen, Dilgren, Chiniya, Griffith, Ding, and Chen}]{shen2025secrepobench}
Chihao Shen, Connor Dilgren, Purva Chiniya, Luke Griffith, Yu~Ding, and Yizheng Chen. 2025.
\newblock Secrepobench: Benchmarking code agents for secure code completion in real-world repositories.
\newblock \emph{arXiv preprint arXiv:2504.21205}.

\bibitem[{Team et~al.(2026)Team, Bai, Bai, Bao, Cai, Cao, Charles, Che, Chen, Chen, Chen, Chen, Chen, Chen, Chen, Chen, Chen, Chen, Chen, Chen, Chen, Chen, Chen, Chen, Chen, Chen, Chen, Chen, Chen, Cheng, Chu, Cui, Deng, Diao, Ding, Dong, Dong, Dong, Dong, Du, Du, Du, Du, Du, Fan, Fang, Feng, Feng, Fu, Fu, Gao, Gao, Ge, Geng, Gong, Gong, Gongque, Gu, Gu, Gu, Guan, Guo, Hao, He, He, He, Hong, Hu, Hu, Hu, Hu, Huang, Huang, Huang, Huang, Jiang, Jiang, Jin, Jing, Lai, Li, Li, Li, Li, Li, Li, Li, Li, Li, Li, Li, Li, Li, Li, Li, Li, Li, Li, Li, Li, Li, Li, Li, Liao, Lin, Lin, Lin, Lin, Liu, Liu, Liu, Liu, Liu, Liu, Liu, Liu, Liu, Liu, Liu, Liu, Liu, Liu, Liu, Liu, Liu, Liu, Lu, Lu, Lu, Luo, Luo, Luo, Ma, Ma, Mao, Mei, Men, Meng, Meng, Miao, Ni, Ouyang, Pan, Pang, Qian, Qin, Qin, Qiu, Qu, Shang, Shao, Shen, Shen, Shi, Shi, Shi, Song, Song, Song, Song, Su, Su, Su, Sui, Sun, Sun, Sun, Sung, Tai, Tang, Tang, Tang, Tang, Tao, Teng, Tian, Tian, Wang, Wang, Wang, Wang, Wang, Wang, Wang, Wang, Wang, Wang, Wang, Wang, Wang,
  Wang, Wang, Wang, Wang, Wang, Wang, Wang, Wang, Wang, Wang, Wang, Wang, Wang, Wang, Wang, Wang, Wang, Wang, Wang, Wang, Wang, Wang, Wang, Wang, Wang, Wei, Wei, Wen, Wen, Wu, Wu, Wu, Wu, Wu, Wu, Wu, Wu, Wu, Xiao, Xie, Xie, Xie, Xin, Xing, Xu, Xu, Xu, Xu, Xu, Xu, Xu, Xu, Xu, Xu, Xu, Xu, Xu, Xu, Xu, Yan, Yan, Yang, Yang, Yang, Yang, Yang, Yang, Yang, Yang, Yang, Yang, Yang, Yang, Yang, Yang, Yao, Ye, Ye, Ye, Yin, Yu, Yu, Yu, Yu, Yuan, Yuan, Yuan, Yue, Zeng, Zha, Zhan, Zhang, Zhang, Zhang, Zhang, Zhang, Zhang, Zhang, Zhang, Zhang, Zhang, Zhang, Zhang, Zhang, Zhang, Zhang, Zhang, Zhang, Zhang, Zhao, Zhao, Zhao, Zhao, Zhao, Zhao, Zhao, Zheng, Zheng, Zheng, Zheng, Zhong, Zhong, Zhong, Zhou, Zhou, Zhou, Zhou, Zhu, Zhu, Zhu, Zhu, Zhu, Zhuang, Zhuang, Zou, and Zu}]{kimiteam2026kimik25visualagentic}
Kimi Team, Tongtong Bai, Yifan Bai, Yiping Bao, S.~H. Cai, Yuan Cao, Y.~Charles, H.~S. Che, Cheng Chen, Guanduo Chen, Huarong Chen, Jia Chen, Jiahao Chen, Jianlong Chen, Jun Chen, Kefan Chen, Liang Chen, Ruijue Chen, Xinhao Chen, and 307 others. 2026.
\newblock \href {https://arxiv.org/abs/2602.02276} {Kimi k2.5: Visual agentic intelligence}.
\newblock \emph{Preprint}, arXiv:2602.02276.

\bibitem[{Team(2026)}]{qwen35blog}
Qwen Team. 2026.
\newblock \href {https://qwen.ai/blog?id=qwen3.5} {Qwen3.5: Accelerating productivity with native multimodal agents}.

\bibitem[{Tripathi and Shah(2025)}]{tripathi2025fundamentals}
Nishith~D Tripathi and Vijay~K Shah. 2025.
\newblock \emph{Fundamentals of O-RAN}.
\newblock John Wiley \& Sons.

\bibitem[{Wadhwa et~al.(2024)Wadhwa, Sonwane, Arora, Mehrotra, Utpala, Bairi, Kanade, and Natarajan}]{wadhwa2024masai}
Nalin Wadhwa, Atharv Sonwane, Daman Arora, Abhav Mehrotra, Saiteja Utpala, Ramakrishna~B Bairi, Aditya Kanade, and Nagarajan Natarajan. 2024.
\newblock Masai: Modular architecture for software-engineering ai agents.
\newblock In \emph{NeurIPS 2024 Workshop on Open-World Agents}.

\bibitem[{Wang et~al.(2025)Wang, Li, Song, Xu, Tang, Zhuge, Pan, Song, Li, Singh, Tran, Li, Ma, Zheng, Qian, Shao, Muennighoff, Zhang, Hui, Lin, Brennan, Peng, Ji, and Neubig}]{wang2025openhands}
Xingyao Wang, Boxuan Li, Yufan Song, Frank~F. Xu, Xiangru Tang, Mingchen Zhuge, Jiayi Pan, Yueqi Song, Bowen Li, Jaskirat Singh, Hoang~H. Tran, Fuqiang Li, Ren Ma, Mingzhang Zheng, Bill Qian, Yanjun Shao, Niklas Muennighoff, Yizhe Zhang, Binyuan Hui, and 5 others. 2025.
\newblock \href {https://openreview.net/forum?id=OJd3ayDDoF} {Openhands: An open platform for {AI} software developers as generalist agents}.
\newblock In \emph{The Thirteenth International Conference on Learning Representations}.

\bibitem[{Weitzel et~al.(2025)Weitzel, Graves, Albin, Zhu, Wuerthwein, Tatineni, Mishin, Khoda, Sada, Smarr, DeFanti, and Graham}]{nrp}
Derek Weitzel, Ashton Graves, Sam Albin, Huijun Zhu, Frank Wuerthwein, Mahidhar Tatineni, Dmitry Mishin, Elham Khoda, Mohammad Sada, Larry Smarr, Thomas DeFanti, and John Graham. 2025.
\newblock \href {https://doi.org/10.1145/3708035.3736060} {The national research platform: Stretched, multi-tenant, scientific kubernetes cluster}.
\newblock In \emph{Practice and Experience in Advanced Research Computing 2025: The Power of Collaboration}, PEARC '25, New York, NY, USA. Association for Computing Machinery.

\bibitem[{Yang et~al.(2024{\natexlab{a}})Yang, Zhang, Yang, Jin, Zhang, Peng, Deng, Miao, Liu, Cui et~al.}]{yang2024execrepobench}
Jian Yang, Jiajun Zhang, Jiaxi Yang, Ke~Jin, Lei Zhang, Qiyao Peng, Ken Deng, Yibo Miao, Tianyu Liu, Zeyu Cui, and 1 others. 2024{\natexlab{a}}.
\newblock Execrepobench: Multi-level executable code completion evaluation.
\newblock \emph{arXiv preprint arXiv:2412.11990}.

\bibitem[{Yang et~al.(2024{\natexlab{b}})Yang, Jimenez, Zhang, Lieret, Yang, Wu, Press, Muennighoff, Synnaeve, Narasimhan et~al.}]{yang2024swe}
John Yang, Carlos~E Jimenez, Alex~L Zhang, Kilian Lieret, Joyce Yang, Xindi Wu, Ori Press, Niklas Muennighoff, Gabriel Synnaeve, Karthik~R Narasimhan, and 1 others. 2024{\natexlab{b}}.
\newblock Swe-bench multimodal: Do ai systems generalize to visual software domains?
\newblock \emph{arXiv preprint arXiv:2410.03859}.

\bibitem[{Yang et~al.(2026)Yang, Morgan, Hiltgen, MacDonald, Gross, Devine, Sareen, Williams, Rifkin, Mizerany, {Michael}, H, {Josh}, {royjhan}, {frob}, {Jeremy}, {Grace}, {nicole pardal}, Madsen, {湛露先生}, Goodhart, {Sam}, ARCHITECT, Ward, Braza, M{\" u}ller, Martinez, Ashimine, Baek, and {slouffka}}]{Yang2026ollama}
Michael Yang, Jeffrey Morgan, Daniel Hiltgen, Bruce MacDonald, Jesse Gross, Patrick Devine, Parth Sareen, Matt Williams, Devon Rifkin, Blake Mizerany, {Michael}, Eva H, {Josh}, {royjhan}, {frob}, {Jeremy}, {Grace}, {nicole pardal}, Dane Madsen, and 11 others. 2026.
\newblock \href {https://github.com/ollama/ollama} {ollama/ollama}.
\newblock https://github.com/ollama/ollama.

\bibitem[{Zhang et~al.(2024)Zhang, Yao, Liu, Feng, Liu, Murthy, Lan, Li, Lou, Xu et~al.}]{zhang2024diversity}
Kexun Zhang, Weiran Yao, Zuxin Liu, Yihao Feng, Zhiwei Liu, Rithesh Murthy, Tian Lan, Lei Li, Renze Lou, Jiacheng Xu, and 1 others. 2024.
\newblock Diversity empowers intelligence: Integrating expertise of software engineering agents.
\newblock \emph{arXiv preprint arXiv:2408.07060}.

\bibitem[{Zhang et~al.(2025)Zhang, He, Zhang, Kang, Li, Xie, Wang, Wang, Huang, Fu et~al.}]{zhang2025swe}
Linghao Zhang, Shilin He, Chaoyun Zhang, Yu~Kang, Bowen Li, Chengxing Xie, Junhao Wang, Maoquan Wang, Yufan Huang, Shengyu Fu, and 1 others. 2025.
\newblock Swe-bench goes live!
\newblock \emph{arXiv preprint arXiv:2505.23419}.

\bibitem[{Zhang et~al.(2026)Zhang, Fang, Xie, Zhang, Yu, Sun, Yang, and Chen}]{zhang2026survey}
Quanjun Zhang, Chunrong Fang, Yang Xie, Yaxin Zhang, Shengcheng Yu, Weisong Sun, Yun Yang, and Zhenyu Chen. 2026.
\newblock A survey on large language models for software engineering.
\newblock \emph{Science China Information Sciences}, 69(4):141102.

\bibitem[{Zou et~al.(2025)Zou, Zhao, Tian, Bariah, Bader, Lestable, and Debbah}]{zou2025telecomgpt}
Hang Zou, Qiyang Zhao, Yu~Tian, Lina Bariah, Faouzi Bader, Thierry Lestable, and Merouane Debbah. 2025.
\newblock Telecomgpt: A framework to build telecom-specific large language models.
\newblock \emph{IEEE Transactions on Machine Learning in Communications and Networking}.

\end{thebibliography}
\newpage
\appendix
\section{Timidness}\label{appx:timmid}
In our evaluation, a recurring behavioral failure mode is \textit{timidness} as mentioned in Section \ref{sec:stage1}. Despite successfully deducing the correct files, grasping the required logic, and understanding the necessary modifications, overly aligned models often halt their autonomous execution. Instead of acting decisively to patch the repository, they generate verbose outputs and wait for explicit user permission or manual file uploads despite running in a headless mode. The following log from the \texttt{Kimi} model illustrates this phenomenon perfectly. The model accurately analyzes the commit description and identifies all five required files, yet concludes its reasoning with a complete refusal to proceed autonomously.

\vspace{1em}
\noindent
\fcolorbox{black}{gray!10}{
\begin{minipage}{\dimexpr\linewidth-2\fboxsep-2\fboxrule\relax}
\ttfamily\small
The user wants me to identify which files need to be changed based on the commit description they provided. They want me to stop and wait for them to add the files to the chat before editing.

Let me analyze the commit description:

1. \textbf{lib/asn1/asn1\_utils.cpp} \ldots mentioned as touched \\
2. \textbf{include/srsgnb/asn1/asn1\_utils.h} \ldots mentioned as touched (contains unbounded\_octstring definition) \\
3. \textbf{unittests/asn1/asn1\_rrc\_nr\_test.cpp} \ldots mentioned as touched \\
4. \textbf{unittests/asn1/asn1\_utils\_test.cpp} \ldots mentioned as touched \\
5. \textbf{apps/examples/du/du\_example.cpp} \ldots mentioned as touched

The commit describes: \\
\ldots Redefining unbounded\_octstring as an extension of byte\_buffer \\
\ldots Introducing octet\_string\_helper namespace with functions (to\_uint, to\_octet\_string, to\_hex\_string, hex\_string\_to\_octets, append\_hex\_string) \\
\ldots Migrating fixed/bounded/unbounded octet string code to use helpers \\
\ldots unbounded\_octstring new APIs: to\_string, from\_string, to\_number, from\_number \\
\ldots pack/unpack logic changes \\
\ldots Updates to tests and usage (copy() calls, bit references)

So the files most likely to need changes are: \\
1. include/srsgnb/asn1/asn1\_utils.h (contains unbounded\_octstring class definition) \\
2. lib/asn1/asn1\_utils.cpp (implementation of helpers and methods) \\
3. unittests/asn1/asn1\_utils\_test.cpp (tests for the changes) \\
4. unittests/asn1/asn1\_rrc\_nr\_test.cpp (tests using octet strings) \\
5. apps/examples/du/du\_example.cpp (usage example)

\vspace{0.5em}
\noindent
\colorbox{yellow!50}{\makebox[\dimexpr\linewidth-2\fboxsep\relax][l]{I should list these and stop....}}
\end{minipage}
}
\vspace{1em}


\section{Standard LLM-as-a-Judge}\label{appx:judge}
We initially experimented with a single-shot setting, which is the standard way to implement LLM judges in the literature \cite{gu2024survey}. Here, the judge received the diffs and the questions simultaneously to produce an output. However, this approach failed because the models either experienced context overload and generated unparsable outputs, or the total input size was so large that it exhausted the manageable context length of our compute setup. After testing several models that we could run locally, the gemma4 31b model was ultimately selected. We then transitioned to a two-tier approach where the TeleJudge framework could maintain context across different models and deliver accurate verdicts 64.05\% of the time, achieving up to 96.9\% for specific models. To further enhance this framework, we explored a third tier aimed at evaluating semantics across files to reduce the overall context length. This adjustment, unfortunately, led to inaccurate results and a loss of context because the native code was abstracted away, prompting us to exclusively adopt the two-tier system for TeleJudge.

\section{Prompts}\label{appx:prompts}

The four main prompt templates utilized for the TeleSWEBench are provided in this section. Specifically, the \textit{Generator} and \textit{Validator} prompt templates pertaining to benchmark generation are available at P1 and P2, respectively. Furthermore, the TeleJudge prompt templates, as explained in Section \ref{sec:evaluation_framework} that are used to evaluate the \textit{File-Level} verdicts and the \textit{Holistic} verdicts are provided in P3 and P4. Please note that the specific \texttt{{difficulty\_requirements}} referenced in the Generator template are detailed in Section~\ref{benchg}.

\begin{tcolorbox}[colback=green!5!white, colframe=green!50!black, title=P1: \textbf{Generator}, fonttitle=\bfseries, arc=4mm]
\scriptsize 
You are an expert software-engineer assistant. Your task is to write a \{difficulty\_description\} instruction for another AI assistant to reproduce a code change.

\vspace{1em}
\noindent Analyze the provided commit information and generate the output according to the selected difficulty.

\vspace{1em}
\noindent\textbf{Commit Information:}\\
\{commit\_json\}

\vspace{1em}
\noindent\textbf{Requirements (\{difficulty\}):}\\
\{difficulty\_requirements\}

\vspace{1em}
\noindent\textbf{Output format:}\\
Generate \{output\_length\}.\\
Use \{output\_style\}.\\
Do not output JSON or code blocks unless explicitly requested.

\end{tcolorbox}
\begin{tcolorbox}[colback=gray!5!white, colframe=blue!60!black, title=P2: \textbf{Validator}, fonttitle=\bfseries, arc=4mm]
\scriptsize 
You are validating a question for GitHub Copilot evaluation.
\noindent\textbf{Commit:} \{commit\_info\}\\
\textbf{Question:} \{generated\_question\}\\
\textbf{Difficulty:} \{difficulty\}
\vspace{1em}

\noindent\textbf{DIFFICULTY LEVEL DEFINITIONS:}

\vspace{0.5em}

\textbf{EASY Difficulty:}\\
Must provide EXACT file paths and line numbers\\
Must specify EXACT code to remove and EXACT code to add\\
Should be self-contained with ALL necessary information
\\
\noindent\textbf{MEDIUM Difficulty:}\\ 
Should explain the "what" and "why" of the change\\
Should mention affected files/areas but NOT exact line numbers\\
Should provide key information (values, function names, etc.)
\\
\noindent\textbf{HARD Difficulty:}\\
Should describe high-level objectives and goals\\
Should NOT mention specific file paths or function names\\
Should provide minimal hints (version numbers, API endpoints, etc.)

\vspace{1em}
\noindent\textbf{VALIDATION CRITERIA:}\\
Is the question clear and unambiguous?\\
Does the question contain appropriate information level for the difficulty?\\
Is the question realistic for GitHub Copilot to answer?\\
Does the question match the actual commit changes?\\
Is the question free from errors or inconsistencies?\\

\noindent Respond with JSON only in this exact format:
\begin{verbatim}
{
    "is_valid": true/false,
    "confidence": 0.0-1.0,
    "reasoning": "Brief explanation"
}
\end{verbatim}

\noindent Where:\\
\texttt{is\_valid}: true if the question is valid, false otherwise\\
\texttt{confidence}: Your confidence level (0.0 to 1.0) that this question is valid\\
\texttt{reasoning}: Brief explanation of your decision

\end{tcolorbox}
\begin{tcolorbox}[colback=red!5!white, colframe=red!60!black, title=P3: \textbf{File-Level Judge}, fonttitle=\bfseries, arc=4mm]
\scriptsize 
You are an expert judge evaluating whether an AI coding assistant correctly addressed a software engineering task on the \textbf{srsRAN Project} 5G codebase (C++, telecom / 3GPP protocols).

\noindent\textbf{You will receive:}
\begin{enumerate}
    \item The original task description (question).
    \item The ground truth fix (the accepted human commit diff) for ONE file.
    \item The copilot's proposed changes for the SAME file.
\end{enumerate}

\noindent\textbf{Accept} if the copilot's changes are functionally equivalent to the ground truth for this file — they address the same root cause or implement the same feature, even if variable names, formatting, ordering, or minor stylistic details differ.

\noindent\textbf{Reject} if the copilot's changes miss the core issue, modify the wrong logic, introduce regressions, or are substantively incomplete.

\noindent\textbf{Return ONLY a JSON object with exactly these fields (no other text):}
\begin{verbatim}
{
    "verdict": "accept" or "reject", 
    "confidence": 0-100, 
    "reasons": ["...", "..."]
}
\end{verbatim}

\noindent \textbf{confidence:} 0 = pure guess, 100 = absolutely certain.\\
\textbf{Keep each reason to at most 25 words.}
\vspace{5pt}
\hrule
\vspace{5pt}
\noindent\textbf{Task description:}\\
\{question\}

\noindent\textbf{Ground truth change for this file:}\\
\{gt\_diff\}

\noindent\textbf{Copilot change for this file:}\\
\{copilot\_diff\}
\noindent\textbf{Return ONLY the JSON verdict.}

\end{tcolorbox}

\begin{tcolorbox}[colback=purple!5!white, colframe=purple!60!black, title=P4: \textbf{Holistic Meta-Judge Prompt}, fonttitle=\bfseries, arc=4mm]
\scriptsize 
You are a senior reviewer making a final decision on whether an AI copilot's commit correctly addresses a task on the srsRAN 5G codebase.
\noindent\textbf{Below you will receive:}
\begin{enumerate}
    \item The original task description.
    \item Per-file verdicts from a first-pass review, each with reasons.
\end{enumerate}
\noindent Read the reasons carefully, think about the commit holistically, and make your own judgment. A single rejected file does NOT automatically mean the whole commit fails — consider whether the rejected file is critical to the task or a minor ancillary change.
\noindent\textbf{Return ONLY a JSON object with exactly these fields (no other text):}
\begin{verbatim}
{
    "verdict": "accept" or "reject", 
    "confidence": 0-100, 
    "reasons": ["...", "..."]
}
\end{verbatim}

\noindent \textbf{confidence:} 0 = pure guess, 100 = absolutely certain.\\
\textbf{Keep each reason to at most 25 words.}
\vspace{5pt}
\hrule
\vspace{5pt}

\noindent\textbf{Task description:}\\
\{question\}

\noindent\textbf{Per-file verdicts:}\\
\{formatted\_file\_verdicts\}

\noindent\textbf{Think holistically and return ONLY the JSON verdict.}

\end{tcolorbox}
\section{Sample Hierarchical Judge Outputs}\label{appx:samples_j}
\begin{tcolorbox}[
    enhanced,
    colback=green!5!white,
    colframe=green!65!black,
    title={Commit: \texttt{7fabbe7} \hfill  Verdict: \textbf{Accept}},
    fonttitle=\bfseries,
    boxrule=1.5pt,
    arc=4mm,
    drop shadow=gray!50!white
]
\scriptsize

\textbf{Assessment Metadata:}
\begin{itemize}
    \setlength{\itemsep}{0pt}
    \item \textbf{Commit SHA:} 

    \texttt{7fabbe7b87a800932a0c74f539cb6d888fed56f1}
    \item \textbf{Question ID:} \texttt{7fabbe7\_medium}
    \item \textbf{Difficulty:} Medium
    \item \textbf{Copilot Model:} \texttt{kimi}
    \item \textbf{Judge Model:} \texttt{gemma4:31b}
    \item \textbf{Confidence:} 100\%
    \item \textbf{Timestamp:} 2026-04-16 03:33:48
\end{itemize}

\tcblower
\scriptsize

\textbf{Overall Reasons:} \\
Implementation of configurable search iterations is correct.\\
Unit tests were expanded as requested with appropriate ranges and assertions.\\
Both files were reviewed and accepted as functionally correct.

\vspace{2mm}
\textbf{File Verdicts:}

\begin{tcolorbox}[
    colback=white,
    colframe=gray!50!black,
    title={\small \texttt{lib/scheduler/support/...}},
    fonttitle=\bfseries,
    boxrule=1pt,
    arc=2mm,
    top=2mm, bottom=2mm
]
\scriptsize

    \textbf{Verdict:} Accept (100\% confidence)\\[2mm]
    \textbf{Reasons:}\\
    The copilot correctly implemented the configurable \texttt{max\_prb\_inc\_iterations} parameter with the correct default value.\\
    The copilot added the requested clarifying comment regarding the TBS derivation approximation, although in a slightly different location than the ground truth.
\end{tcolorbox}

\begin{tcolorbox}[
    colback=white,
    colframe=gray!50!black,
    title={\small \texttt{unittests/scheduler/...}},
    fonttitle=\bfseries,
    boxrule=1pt,
    arc=2mm,
    top=2mm, bottom=2mm
]
\scriptsize

    \textbf{Verdict:} Accept (100\% confidence)\\[2mm]
    \textbf{Reasons:} \\
    The copilot's changes are functionally identical to the ground truth, including the new TBS ranges, steps, and assertion logic.\\
    The only difference is the renaming of 'long' to 'large' for the TBS variable, which is a stylistic choice.
\end{tcolorbox}

\end{tcolorbox}

\begin{figure*}[t]
\centering
\begin{tcolorbox}[
  width=\textwidth,
  colback=gtbg,colframe=black!75,
  title=\textbf{Commit \textit{70c8b9d}: Ground Truth},
  fonttitle=\bfseries
]
\footnotesize
\textbf{Commit intent:} remove DCI dependency on \texttt{vrb\_to\_prb::mapping\_type} and use a boolean
\texttt{interleaved\_vrb\_prb\_mapping} across packing/building/tests.\\[1mm]

\textbf{Files changed (6):}

\texttt{include/srsran/ran/pdcch/dci\_packing.h},

\texttt{lib/ran/pdcch/dci\_packing.cpp},

\texttt{lib/scheduler/support/dci\_builder.cpp},

\texttt{lib/scheduler/support/sch\_pdu\_builder.cpp},

\texttt{tests/unittests/fapi\_adaptor/mac/messages/helpers.cpp},

\texttt{tests/unittests/ran/pdcch/dci\_packing\_test.cpp}.

\vspace{1.5mm}
\hrule
\vspace{2mm}

\ttfamily\scriptsize
\textbf{include/srsran/ran/pdcch/dci\_packing.h}\\
\textcolor{red}{- \#include "srsran/ran/resource\_allocation/vrb\_to\_prb.h"}\\
\textcolor{red}{- vrb\_to\_prb::mapping\_type vrb\_to\_prb\_mapping;}\\
\textcolor{green!60!black}{+ bool interleaved\_vrb\_prb\_mapping;}\\
\normalfont\scriptsize $\ldots$ (same replacement across c\_rnti/p\_rnti/si\_rnti/ra\_rnti/tc\_rnti structs)

\vspace{1.5mm}
\ttfamily\scriptsize
\textbf{lib/ran/pdcch/dci\_packing.cpp}\\
\textcolor{red}{- payload.push\_back(config.vrb\_to\_prb\_mapping != vrb\_to\_prb::mapping\_type::non\_interleaved, 1);}\\
\textcolor{green!60!black}{+ payload.push\_back(config.interleaved\_vrb\_prb\_mapping, 1);}\\
\normalfont\scriptsize $\ldots$ (5 packing paths updated)

\vspace{1.5mm}
\ttfamily\scriptsize
\textbf{lib/scheduler/support/dci\_builder.cpp}\\
\textcolor{red}{- si\_dci.vrb\_to\_prb\_mapping = vrb\_to\_prb::mapping\_type::non\_interleaved;}\\
\textcolor{green!60!black}{+ si\_dci.interleaved\_vrb\_prb\_mapping = false;}\\
\normalfont\scriptsize $\ldots$ (p\_rnti, ra\_rnti, tc\_rnti, c\_rnti builders aligned)

\vspace{1.5mm}
\ttfamily\scriptsize
\textbf{lib/scheduler/support/sch\_pdu\_builder.cpp}\\
\textcolor{red}{- pdsch.vrb\_prb\_mapping = dci\_cfg.vrb\_to\_prb\_mapping;}\\
\textcolor{green!60!black}{+ pdsch.vrb\_prb\_mapping = vrb\_to\_prb::mapping\_type::non\_interleaved;}\\
\normalfont\scriptsize $\ldots$ (5 call sites)

\vspace{1.5mm}
\ttfamily\scriptsize
\textbf{tests/.../helpers.cpp}\\
\textcolor{red}{- return \{1,2,3, vrb\_to\_prb::mapping\_type::non\_interleaved, 0,2,1\};}\\
\textcolor{green!60!black}{+ return \{1,2,3, false, 0,2,1\};}

\vspace{1.5mm}
\ttfamily\scriptsize
\textbf{tests/.../dci\_packing\_test.cpp}\\
\textcolor{red}{- config.vrb\_to\_prb\_mapping = ... ? interleaved\_n2 : non\_interleaved;}\\
\textcolor{green!60!black}{+ config.interleaved\_vrb\_prb\_mapping = vrb\_to\_prb\_mapping\_dist(rgen) > 0;}\\
\textcolor{red}{- expected.push\_back(config.vrb\_to\_prb\_mapping != ...non\_interleaved);}\\
\textcolor{green!60!black}{+ expected.push\_back(config.interleaved\_vrb\_prb\_mapping);}
\end{tcolorbox}
\vspace{1.5mm}

\begin{tcolorbox}[colback=white,colframe=easyc,title=\textbf{Easy}]
\footnotesize
\textbf{Question:}
Modify the DCI path to remove enum-based VRB-to-PRB mapping usage and replace it with a boolean flag.
In \texttt{include/srsran/ran/pdcch/dci\_packing.h}, replace each
\texttt{vrb\_to\_prb::mapping\_type vrb\_to\_prb\_mapping;} with
\texttt{bool interleaved\_vrb\_prb\_mapping;} (across all relevant DCI config structs), and remove the obsolete include of
\texttt{vrb\_to\_prb.h}. In \texttt{lib/ran/pdcch/dci\_packing.cpp}, replace each mapping-bit emission that \ldots Then update scheduler builders and unit tests/helpers so values are assigned and validated as booleans (typically \texttt{false} for default non-interleaved) \ldots
\end{tcolorbox}

\begin{tcolorbox}[colback=white,colframe=medc,title=\textbf{Medium}]
\footnotesize
\textbf{Question:}
Refactor the DCI fallback/control flow so VRB-to-PRB interleaving is represented as a lightweight boolean instead of the previous enum dependency.
Apply the migration consistently across
(1) DCI configuration structs, \ldots
(3) scheduler-side DCI construction, and
(4) tests/helpers that synthesize and check payload bits.
Preserve functional semantics: previous ``non-interleaved'' behavior should remain \ldots The expected end state is that enum-specific plumbing disappears from this path and all affected tests validate the boolean representation end-to-end \ldots
\end{tcolorbox}

\begin{tcolorbox}[colback=white,colframe=hardc,title=\textbf{Hard}]
\footnotesize
\textbf{Question:}
Consolidate and simplify the DCI packing surface by removing the direct dependency on
\texttt{vrb\_to\_prb::mapping\_type} while keeping externally observable behavior unchanged.
The migration must be coherent across headers, packers, scheduler builders, and verification utilities, and should avoid partial conversions that leave mixed representations \ldots
When done, the mapping indicator should be encoded from a single boolean field and construction paths should keep a clear non-interleaved default unless explicitly overridden.
\textbf{Hint:} propagate a boolean (e.g., \texttt{interleaved\_vrb\_prb\_mapping}) through all encode/validate paths, including helper-generated fixtures \ldots
\end{tcolorbox}\caption{One \texttt{srsRAN\_Project} ground-truth diff (commit \texttt{70c8b9d}) shown with representative per-file chunks and corresponding questions across easy, medium and hard.}
\label{fig:shared-gt-70c8b9d}
\end{figure*}

\section{Sample Questions across Difficulty Tiers}\label{appx:samples_q}
A sample from TeleSWEBench with associated questions and ground truth is shown in Figure \ref{fig:shared-gt-70c8b9d}.

\section{Extracting Test Cases}\label{appx:tc}


To evaluate the generated code, we leverage the framework’s native unit-testing hierarchy, which follows a locality principle: test files are typically placed under dedicated test directories whose paths closely mirror the locations of their corresponding implementation files. In addition, test files often share the same base name as the implementation file, commonly with a \_test suffix appended. This structural symmetry between implementation and testing components enables the efficient extraction and isolated evaluation of targeted network modules within our benchmarking framework. To identify implementation–test pairs, we employ a name-matching procedure that compares implementation filenames with candidate test files while accounting for common naming patterns such as suffix additions (e.g., \_test) and path similarities. We further apply a threshold-based matching criterion to retain only sufficiently similar pairs and reduce spurious associations. This process allows us to systematically construct implementation–test mappings across the repository while preserving high confidence in the extracted relationships.

However, test coverage is not available across all repository snapshots. A key challenge is that complete test vectors, together with all associated files required for successful compilation, are only consistently available in merge commits. Intermediate commits frequently contain missing dependencies or incomplete test artifacts that prevent standalone execution. Consequently, our benchmark construction primarily relies on merge commits to ensure that extracted test suites remain compilable and executable in isolation. After establishing implementation–test mappings, the next step is to associate benchmark questions with their corresponding tests. To achieve this, we utilize the ground-truth file modifications available in the benchmark dataset. Specifically, if at least one file modified in the ground-truth patch matches the implementation file associated with a test case, we assign that benchmark question to the corresponding test suite. This strategy ensures that the functionality affected by the target implementation file is validated by its native testing infrastructure.


\end{document}